\documentclass[aps,pra,onecolumn,showpacs,superscriptaddress,floatfix]{revtex4}

\usepackage{dcolumn}
\usepackage{graphicx}
\usepackage{epsf}
\usepackage{amsmath}
\usepackage{amssymb}
\usepackage{bm}
\usepackage{times}

\def\Za{{Z\alpha}}

\def\rms{<\!\!\!\,r^2\!\!\!>^{1/2}}
\newcolumntype{.}{D{x}{}{-1}}

\begin{document}

\title{Two-loop QED corrections in few-electron ions}

\author{V.~A.~Yerokhin}
 \affiliation{Max--Planck--Institut f\"ur Kernphysik, Saupfercheckweg 1, 69117
Heidelberg, Germany}
 \affiliation{Department of Physics, St.~Petersburg State
University, Oulianovskaya 1, St.~Petersburg 198504, Russia}

\author{P.~Indelicato}
\affiliation{Laboratoire Kastler-Brossel, \'Ecole Normale Sup\'erieure et Universit\'e P. et M.
Curie, Case 74, 4 pl.~Jussieu, F-75252, France}

\author{V.~M.~Shabaev}
\affiliation{Department of Physics, St.~Petersburg State University, Oulianovskaya 1,
St.~Petersburg 198504, Russia}

\begin{abstract}
Results of a calculation valid to all orders in the nuclear-strength parameter $\Za$
are presented for the two-loop Lamb shift, notably for the two-loop self-energy
correction, for the ground and first excited states of ions with the nuclear charge
numbers $Z=60$-100. A detailed comparison of the all-order calculation with earlier
investigations based on an expansion in the parameter $\Za$ is given. The
calculation removes the largest theoretical uncertainty for the $2p_j$-$2s$
transition energies in heavy Li-like ions and is important for interpretation of
experimental data.

\end{abstract}

\pacs{31.30.Jv, 31.30.-i, 31.10.+z}

\maketitle

%%%%%%%%%%%%%%%%%%%%%%%%%%%%%%%%%%%%%%%%%%%%

\section{Introduction}

Measurements of the $2p_j$-$2s$ transition energies in heavy Li-like ions
\cite{schweppe:91,beiersdorfer:98,brandau:04,beiersdorfer:05} have lately reached a
fractional accuracy of 0.03\% with respect to the total QED contribution. This
corresponds to a 10\% sensitivity of the experimental results to the two-loop QED
effects. These measurements provide an excellent possibility for identification of
the two-loop Lamb shift and for testing the bound-state QED theory up to second
order in $\alpha$ in the strong-field regime. Adequate interpretation of
experimental data requires theoretical investigations of the two-loop QED effects
valid to all orders in the nuclear-strength parameter $\Za$.

All-order calculations of the two-loop QED effects and, first of all, the two-loop
self-energy correction are important also for low-$Z$ ions because of a very slow
convergence of the corresponding $\Za$ expansion. The higher-order (in $\Za$)
two-loop QED effects presently yield the second largest uncertainty in the
theoretical prediction for the ground-state Lamb shift in hydrogen (after the proton
charge distribution effect). Improved theoretical results for the $1s$ and $2s$ Lamb
shift will be required in the near future for a more precise determination of the
Rydberg constant, when an improved value for the proton charge radius is obtained
from the muonic hydrogen experiment \cite{nebel:thisvolume}.

\begin{figure}
\centerline{
\resizebox{0.6\columnwidth}{!}{%
  \includegraphics{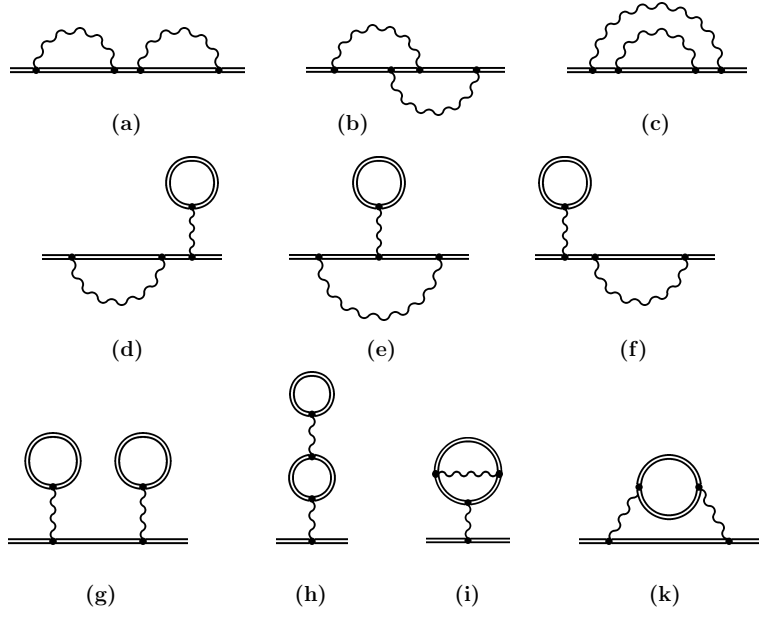}
}}
 \caption{Two-loop one-electron QED corrections. Gauge-invariant subsets are
referred to as SESE (a-c), SEVP (d-f), VPVP (g-i), S(VP)E (k). \label{fig:2order}}
\end{figure}

The complete set of two-loop one-electron QED corrections (also referred to as the
two-loop Lamb shift) is graphically represented in Fig.~\ref{fig:2order}. In this
investigation, we will be mainly concerned with the two-loop self-energy correction
represented by diagrams (a)-(c), which will be evaluated rigorously to all orders in
$\Za$. The other diagrams in Fig.~\ref{fig:2order} will be calculated as well;
diagrams (d)-(g), rigorously and diagrams (h)-(k), within the free-loop
approximation, i.e., keeping the leading term of the expansion of fermion loops in
terms of the binding potential. In the one-loop case, the free-loop approximation
corresponds to the Uehling potential and yields the dominant contribution even for
high-$Z$ ions like uranium. We assume that the free-loop approximation is reasonably
adequate in the two-loop case also.

%%%%%%%%%%%%%%%%%%%%%%%%%%%%%%%%%%%%%%%%%%%%%%%%%%%%%%%%%%%%%%%%%%%%%%%%
%%%%%
%%%%%
%%%%%%%%%%%%%%%%%%%%%%%%%%%%%%%%%%%%%%%%%%%%%%%%%%%%%%%%%%%%%%%%%%%%%%%

%%%%%%%%%%%%%%%%%%%%%%%%%%%%%%%%%%%%%%%%%%%%%%%%%%%%%%%%%%%%%%%%%%%%%%%%%%%%%%%%

\section{Two-loop self-energy}

The two-loop self-energy correction to the energy shift is
conveniently represented in terms of the dimensionless function
$F(\Za)$ defined (in relativistic units $\hbar=c=1$) by
\begin{equation}\label{FalphaZ}
 \delta E  = m\, \left(\frac{\alpha}{\pi}\right)^2\,
                 \frac{(Z\,\alpha)^4}{n^3}\,F(Z\,\alpha)\,,
\end{equation}
where $n$ is the principal quantum number. The $\Za$ expansion of the function $F$
reads
\begin{align} \label{aZexp}
 F(Z\alpha) =& \ B_{40}+
   (Z\alpha)\,B_{50} + (Z\alpha)^2\,
  \bigl[ L^3\, B_{63}
%\nonumber \\ &
    +L^2\, B_{62}
 +  L\,B_{61} + G^{\rm h.o.}(\Za) \bigr]
  \,,
\end{align}
where  $L =\ln[(Z\alpha)^{-2}]$  and $G^{\rm h.o.}$ is a non-perturbative remainder
whose expansion starts with a constant, $G^{\rm h.o.}(\Za) = B_{60}+ \Za
\,(\cdots)\,$.

The leading term in Eq.~(\ref{aZexp}), $B_{40}$, is related to the free-electron
form-factors and is known for a long time. Its first complete evaluation was
reported in \cite{appelquist:70}. The next term $B_{50}$ was calculated only
relatively recently by Pachucki \cite{pachucki:94} and by Eides and Shelyuto
\cite{eides:95:pra}. The result for the leading logarithmic contribution, $B_{63}$,
was first reported in \cite{karshenboim:93:jetp}. A considerable discussion about
the correctness of the method of derivation followed this publication (see
\cite{yerokhin:01:prl} and references therein). Finally, this result was rigorously
re-derived by Pachucki in \cite{pachucki:01:pra}. In that work, Pachucki also
derived the remaining logarithmic terms for $ns$ states, $B_{62}(ns)$ and
$B_{61}(ns)$. An additional contribution to the coefficient $B_{61}(1s)$ was
recently identified and evaluated in \cite{czarnecki:05}. The coefficient
$B_{62}$ for $np$ states was calculated in \cite{karshenboim:96:jpb}. The values of
$B_{61}$ for states with $l\ge1$ as well as the differences $B_{60}(ns) -
B_{60}(1s)$ and $B_{60}(np_j)-B_{60}(np_j)$ were recently presented in
\cite{czarnecki:05}. There are no complete results available at present for
the coefficient $B_{60}$ for single states. However, its non-relativistic part $b_L$
(also termed as the two-loop Bethe logarithm) was calculated for $1s$ and $2s$
states in \cite{pachucki:03:prl} and later for higher excited states in
\cite{jentschura:04:B60,jentschura:thisvolume}. This part presumably yields the
dominant contribution to $B_{60}$; the uncertainty due to uncalculated terms was
estimated in \cite{pachucki:03:prl} to be 15\% for the $1s$ and $2s$ states.

The summary of the results available for the $\Za$-expansion coefficients of the
two-loop self-energy correction reads
\begin{eqnarray} \label{coef}
  B_{40} &=& \left[-\frac{163}{72}-\frac{85}{216}\,\pi^2
   + \frac32 \, \pi^2 \ln 2\, -\frac94\, \zeta(3)  \right]\delta_{l0}
\nonumber \\
&& {} -  \left[ -\frac{31}{16}  + \frac{5}{12}\,{\pi}^2 -
  \frac12 {\pi }^2\,\ln 2 + \frac34\,\zeta(3) \right]
    \frac{1-\delta_{l0}}{\kappa\,(2\,l+1)}\,,
 \\
  B_{50} &=& -24.2668\,(31)\,\delta_{l0}\,,
 \\
  B_{63} &=& -\frac{8}{27}\,\delta_{l0}\,,
 \\
  B_{62}(ns) &=& \frac{16}{9}\,\left(
  \frac{13}{12}-\ln 2
     +\frac1{4n^2}-\frac1n-\ln n
      + \psi(n)+C \right)\,,
 \\
  B_{62}(np) &=& \frac{4}{27}\,\frac{n^2-1}{n^2}\,,
 \\
  B_{62}(nd) &=& 0\,,
\\
  B_{61}(ns) &=&
  \frac{15473}{2592} + \frac{1039}{432}\,{\pi }^2
  - \frac{152}{27}\,\ln 2 -
  \frac{2}{3}\,{\pi }^2\,\ln 2 + \frac{40}{9}\,\ln^2 2 +
   \zeta (3) + \frac{4}{3}\,N(ns)
\nonumber \\ &&
  {} + \left( \frac{80}{27}-\frac{32}{9}\,\ln 2 \right)
   \left( \frac34 +\frac1{4n^2}-\frac1n -\ln n +\psi(n)+C
   \right)\,,
 \\
  B_{61}(np_{1/2}) &=& \frac43\,N(np)+ \frac{n^2-1}{n^2}
    \left(\frac{38}{81}-\frac{8}{27}\,\ln 2\right) \,,
   \label{eqB61a}
 \\
  B_{61}(np_{3/2}) &=& \frac43\,N(np)+ \frac{n^2-1}{n^2}
    \left(\frac{11}{81}-\frac{8}{27}\,\ln 2\right) \,,
   \label{eqB61b}
 \\
  B_{61}(nd) &=& 0\,,
 \end{eqnarray}
 \begin{eqnarray}
  B_{60}(ns) &=& b_L(ns)+ \frac{10}{9}\,N(ns) + \ldots\,,
   \label{eqB60s}
 \\
  B_{60}(np) &=& b_L(np) + \ldots\,,
   \label{eqB60p}
\end{eqnarray}
where $\kappa$ is the Dirac angular-momentum quantum number, $l$ is the orbital
quantum number, $l = |\kappa+1/2|-1/2$, $\zeta(n)$ is the Riemann zeta function,
$\psi(n)$ is the logarithmic derivative of the gamma function, $C$ is the Euler
constant, and $(\ldots)$ denotes uncalculated terms. Complete results are available
for the differences of the coefficients $B_{60}$ \cite{czarnecki:05}, in
particular,
\begin{eqnarray}
   \label{eqB602s1s}
  B_{60}(2s)-  B_{60}(1s) &=& b_L(2s)-b_L(1s)+ 0.318\,486\,,
 \\
   \label{eqB602p31}
  B_{60}(2p_{3/2})- B_{60}(2p_{1/2}) &=& -0.361\,196\,.
\end{eqnarray}
Accurate numerical values for the function $N(nl)$ were obtained in
\cite{jentschura:03:jpa,czarnecki:05}. For the states with $n=1$ and 2, they
are given by
\begin{eqnarray}
  N(1s) &=& 17.855\,672\,03\,(1)\,, \\
  N(2s) &=& 12.032\,141\,58\,(1)\,, \\
  N(2p) &=& \ \  0.003\,300\,635\,(1)\,.
\end{eqnarray}
The two-loop Bethe logarithms for these states are
\cite{pachucki:03:prl,jentschura:thisvolume}:
\begin{eqnarray}
  b_L(1s) &=& -81.4\,(3)\,, \\
  b_L(2s) &=& -66.6\,(3)\,, \\
  b_L(2p) &=& \ \ -2.2\,(3)\,.
\end{eqnarray}
We note that the formulas (\ref{eqB61a}), (\ref{eqB61b}) for the two-loop
self-energy correction were obtained from the full two-loop results in
\cite{czarnecki:05} by subtracting the contribution due to diagrams with
closed fermion loops, which is \cite{jentschura:priv}
\begin{equation}
    B_{61}(np, {\rm VP}) = -\frac{8}{135}\,\frac{n^2-1}{n^2}\,.
\end{equation}

We now turn to our calculation of the two-loop self-energy correction to all orders
in $\Za$. The starting point of our consideration is the Furry picture, where the
interaction of an electron with the nucleus is taken into account to all orders
right from the beginning. As a consequence of this choice, we have to deal with the
bound-electron propagators, whose structure is much more complicated than that of
the free-electron propagators. A method for the evaluation of the two-loop
self-energy correction was developed for the ground state in our previous studies
\cite{yerokhin:03:prl,yerokhin:05:sese}. In the present
investigation, we extend it to the excited states. The general procedure for
isolation and cancelation of ultraviolet and infrared divergences is similar to that
for the $1s$ state, but the actual calculational scheme requires substantial
modifications due to a more complicated pole and angular-momentum structure of
expressions involved. Details of our calculation are cumbersome and will be
published elsewhere; in this paper we concentrate on presentation and analysis of
the results obtained and discuss their experimental consequences.

%%%%%%%%%%%%%%%%%%%%%%%%%%%%%%%%%%%%%%%%%%%%%%%%%%%%%%%%%%%%%%%%%%%%%%%%
%%%%%
%%%%%
%%%%%%%%%%%%%%%%%%%%%%%%%%%%%%%%%%%%%%%%%%%%%%%%%%%%%%%%%%%%%%%%%%%%%%%
\begin{figure}
\centerline{
\resizebox{\textwidth}{!}{%
  \includegraphics{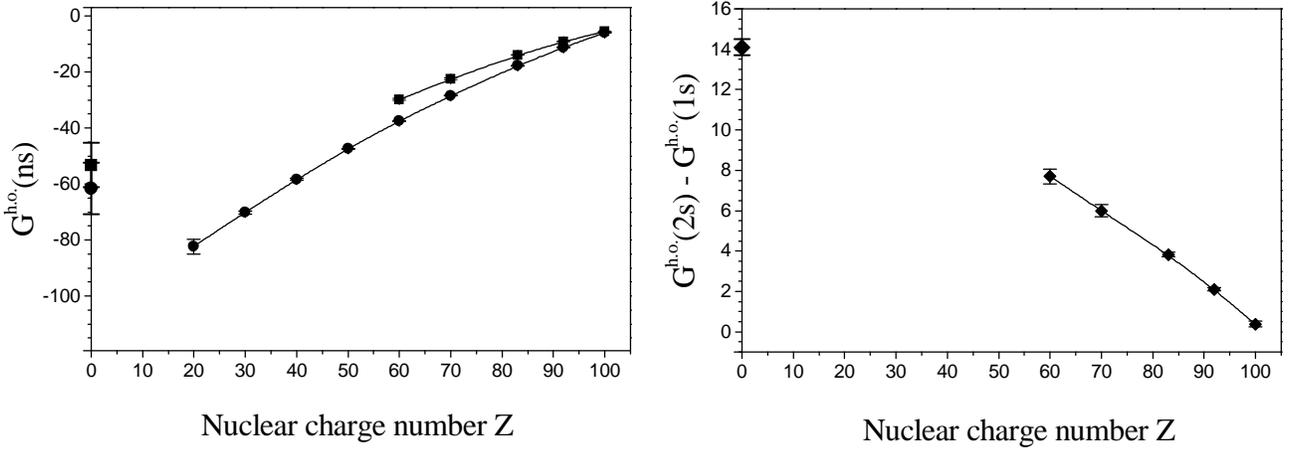}
}}
 \caption{The higher-order remainder function $G^{\rm h.o.}$
for the $1s$ and $2s$ states (dots and squares on the left graph, respectively) and
for the difference $\Delta_s^{\rm h.o.}$ (the right graph).
 \label{fig:ho:ns}}
\end{figure}

\begin{figure}
\centerline{
\resizebox{\textwidth}{!}{%
  \includegraphics{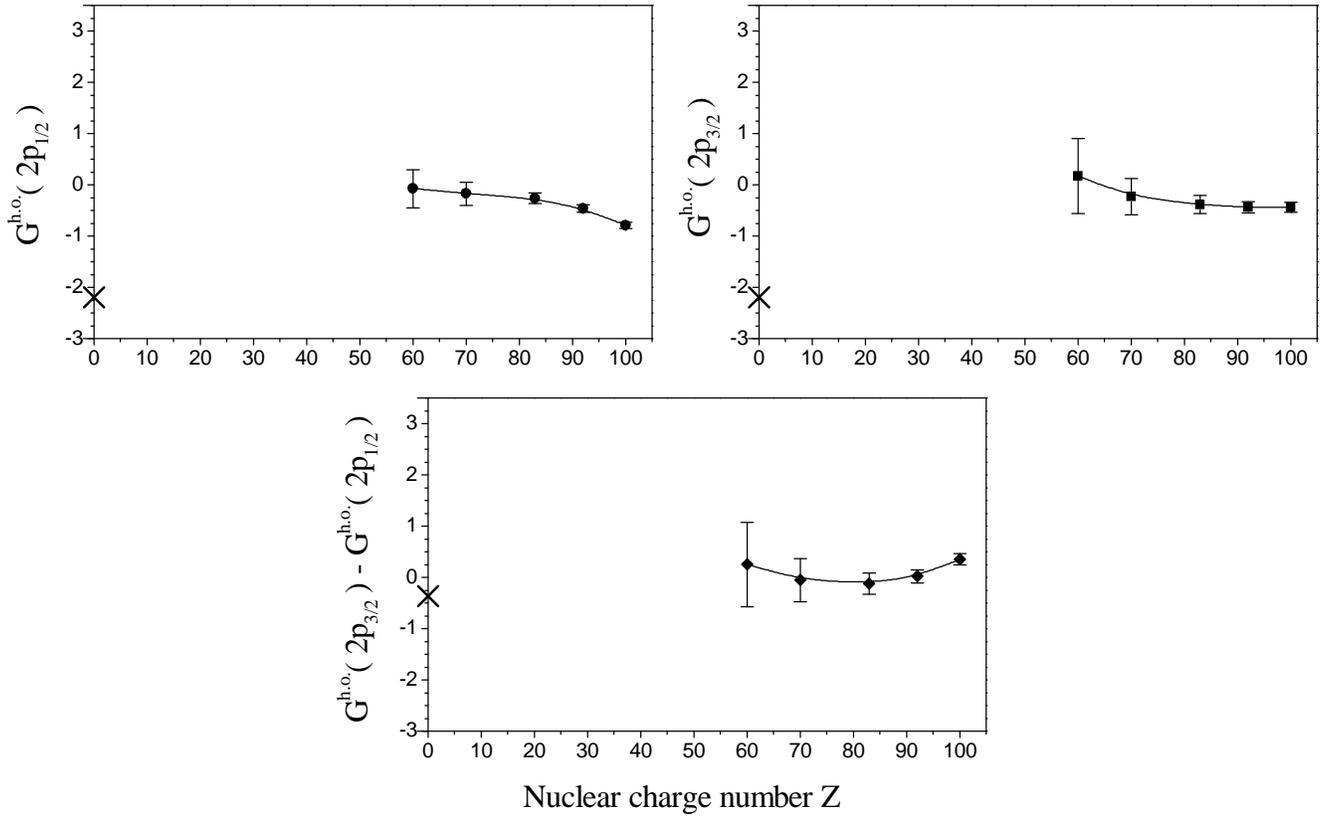}
}}
 \caption{The higher-order remainder function $G^{\rm h.o.}$
for the $2p_{1/2}$ state (the left upper graph), $2p_{2/3}$ state (the right upper
graph), and for the difference $\Delta_p^{\rm h.o.}$ (the lower graph).
 \label{fig:ho:np}}
\end{figure}

We performed our all-order calculations of the two-loop self-energy correction for
the $1s$, $2s$, $2p_{1/2}$, and $2p_{3/2}$ states of ions with $Z=60$, 70, 83, 92
and 100. The results can be conveniently expressed in terms of the higher-order
remainder $G^{\rm h.o.}$ defined by Eq.~(\ref{aZexp}). The values of the remainder
as a function of $Z$ are plotted in Fig.~\ref{fig:ho:ns} for the $1s$ and $2s$
states and in Fig.~\ref{fig:ho:np} for the $2p_{1/2}$ and $2p_{3/2}$ states. The
results for the $1s$ state and $Z < 60$ are taken from our previous investigation
\cite{yerokhin:05:sese} (the points with $Z=10$ and 15 are omitted because of large
numerical uncertainty). We present separate graphs for individual states and for the
differences $\Delta_s^{\rm h.o.} = G^{\rm h.o.}(2s)-G^{\rm h.o.}(1s)$ and
$\Delta_p^{\rm h.o.} =G^{\rm h.o.}(2p_{3/2})-G^{\rm h.o.}(2p_{1/2})$. On the
ordinate axis of the graphs, the limiting values of the higher-order remainder at
$Z=0$ are indicated, as obtained within the $\Za$-expansion approach. The status of
these limiting values is different for single states and for the differences
$\Delta_s^{\rm h.o.}$ and $\Delta_p^{\rm h.o.}$. For the single states, the limiting
values represent incomplete results for the coefficients $B_{60}$ given by
Eqs.~(\ref{eqB60s}) and (\ref{eqB60p}). The error bars indicated for the $1s$ and
$2s$ states correspond to the 15\% uncertainty suggested in \cite{pachucki:03:prl}.
For $2p$ states, the uncertainty is undefined, although it is believed to be
significantly smaller than the two-loop Bethe logarithm. For the differences
$\Delta_s^{\rm h.o.}$ and $\Delta_p^{\rm h.o.}$, the limiting values are known much
better and given by Eqs.~(\ref{eqB602s1s}) and (\ref{eqB602p31}).

%%%%%%%%%%%%%%%%%%%%%%%%%%%%%%%%%%%%%%%%%%%%%%%%%%%%%%%%%%%%%%%%%%%%%%%%
%%%%%
%%%%%
%%%%%%%%%%%%%%%%%%%%%%%%%%%%%%%%%%%%%%%%%%%%%%%%%%%%%%%%%%%%%%%%%%%%%%%

Characterizing the comparison presented in Fig.~\ref{fig:ho:ns}, we observe that our
all-order results do not seem to agree well with the the analytical results to order
$\alpha^2 (\Za)^6$ for $1s$ and $2s$ states separately whereas a rather good
agreement is found for the difference $\Delta_s^{\rm h.o.}$. For the $2p_j$ states
presented in Fig.~\ref{fig:ho:np}, the situation is even less definite, due to
smaller numerical values of the higher-order remainder function. But we again
observe that agreement for the difference $\Delta_p^{\rm h.o.}$ is much better than
for the single states. It should be mentioned that the agreement observed for the
differences is a valuable evidence in favor of reliability of our all-order results.
The reason is that numerical evaluations for different single states are completely
independent and individual contributions, {\it e.g.}, to the functions $F(1s)$ and
$F(2s)$ are even of different orders of magnitude. It is thus very unlikely that a
contribution appears in numerical evaluations that vanishes identically in the
difference, which is contrary to the situation in the $\Za$-expansion calculations.

Assuming correctness of both the all-order and the $\Za$-expansion calculation, we
can surmise two possible explanations of the situation observed. The first
possibility is that uncalculated contributions to the coefficients $B_{60}$ for
single states are larger than previously expected and, when calculated, they will shift
the limiting values $G^{\rm h.o.}(Z=0)$ in Figs.~\ref{fig:ho:ns} and \ref{fig:ho:np}
considerably. The second possibility is that remarkably large logarithmic terms
appear in the $\Za$ expansion to order $\alpha^2(\Za)^7$ and induce a very rapidly
varying structure in the $Z$ dependence of the remainder $G^{\rm h.o.}(Z)$ in the
low-$Z$ region. The both scenarios will have a significant influence on the
theoretical values of the higher-order two-loop QED effects for the $1s$ and $2s$
states in hydrogen.

%%%%%%%%%%%%%%%%%%%%%%%%%%%%%%%%%%%%%%%%%%%%%%%%%%%%%%%

\section{Two-loop Lamb shift in Li-like ions}

The best opportunity for experimental identification of the two-loop QED effects in
the strong binding field is presently offered by measurements of the $2p_j$-$2s$
transition energies in Li-like ions. In this work, we present results of our
calculations of all two-loop corrections depicted in Fig.~\ref{fig:2order} for the
$2p_{3/2}$-$2s$ transition in Bi$^{80+}$ and the $2p_{1/2}$-$2s$ transition in
U$^{89+}$, for which most accurate experimental data are available. Numerical
results for individual subsets of diagrams defined in Fig.~\ref{fig:2order} are
presented in Table~\ref{tab:3} under the entry ``Two-loop QED''. The SESE subset
represents the two-loop self-energy correction, which is the main result of our
investigation. The SEVP(d-f) diagrams and the VPVP(g) diagram were calculated to all
orders in $\Za$ without any approximations involved, whereas the VPVP(h,i) and
S(VP)E diagrams were evaluated within the free-loop approximation, {\it i.e.},
keeping the first nonvanishing contribution in the expansion of the fermion loops in
terms of the binding potential. The error bars specified for these corrections are
estimations of uncertainty due to the approximation employed. They were obtained by
multiplying the contribution of diagrams (h,i) by a factor of $(\Za)^2$ and that of
diagram (k) -- by a factor of $3\,(\Za)$. The factor of $3\,(\Za)$ in the latter
estimation arises as a ratio of the leading-order contribution beyond the free-loop
approximation for the diagram (k), $-0.386\, (\alpha/\pi)^2(\Za)^5$
\cite{pachucki:93:pra}, and the leading-order contribution within this
approximation, $0.142\, (\alpha/\pi)^2(\Za)^4$ \cite{lautrup:70}. The finite nuclear
size effect was taken into account in our evaluation of the diagrams (d)-(i),
whereas the other diagrams were calculated for the point nuclear model. In the case
of uranium, our results for the diagrams with closed fermion loops are in good
agreement with those reported previously
\cite{beier:88,persson:96:pra,mallampalli:96,plunien:98:epj}.

%%%%%%%%%%%%%%%%%%%%%%%%%%%%%%%%%%%%%%%%%%%%%%%%%%%%%%%%%%%%%%%%%%%%%%%%%%%
%
%       2p-2s transition
%
%%%%%%%%%%%%%%%%%%
\begin{table}
\caption{Individual contributions to the $2p_{3/2}$-$2s$ and $2p_{1/2}$-$2s$
transition energies in Li-like bismuth and uranium, in eV.
 \label{tab:3} }
\begin{center}
\begin{tabular}{ll..}
\hline\\
         && \multicolumn{1}{c}{$2p_{3/2}$-$2s$,$\ Z=83$}
                              & \multicolumn{1}{c}{$2p_{1/2}$-$2s$,$\ Z=92$} \\
 \hline\\[-9pt]
Dirac value           &&   2792.x21 \,(3)    &   -33.2x7 \,(9) \\
One-photon exchange   &&     23.x82          &   368.8x3        \\
Two-photon exchange   &&     -1.x61          &   -13.3x7        \\
Three-photon exchange &&     -0.x02 \,(2)    &     0.1x5 \,(7)  \\
One-loop QED          &&    -27.x48          &   -42.9x3        \\
Screened QED          &&      1.x15 \,(4)    &     1.1x6 \,(3)  \\
Two-loop QED & SESE     &     0.x15          &     0.3x0       \\
             & SEVP     &    -0.x10          &    -0.1x9       \\
             &VPVP (g)  &     0.x02          &     0.0x4       \\
             &VPVP (h,i)&     0.x07\,(3)     &     0.1x0\,(5)  \\
             & S(VP)E   &    -0.x01\,(2)     &    -0.0x2\,(5)  \\
Recoil                &&     -0.x07          &    -0.0x7        \\
Nuclear polarization  &&                     &     0.0x4 \,(2)  \\
 \hline\\[-9pt]
Total theory          &&   2788.x12\,(7)    &   280.7x6 \,(14) \\
Experiment            &&   2788.x14\,(4)\, \mbox{\cite{beiersdorfer:98}}
                                           & 280.6x45\,(15)\,   \mbox{\cite{beiersdorfer:05}} \\
                     &&                     & 280.5x16\,(99)\,   \mbox{\cite{brandau:04}} \\
                     &&                     & 280.5x9\,(10)\,   \mbox{\cite{schweppe:91}} \\
\hline\\
\end{tabular}
\end{center}
\end{table}

We now explain the other theoretical contributions to the transition energies
presented in Table~\ref{tab:3}. The entry labeled ``Dirac value'' represents the
transition energies as obtained from the Dirac equation with the nuclear potential
induced by the standard two-parameter Fermi nuclear-charge distribution. Numerical
values for the nuclear-charge root-mean-square (rms) radii were taken from
\cite{angeli:04}, $\rms = 5.851(7)$~Fm for uranium and $5.521(3)$~Fm for bismuth.
The dependence of the Dirac value on the nuclear model was conservatively estimated
by comparing the results obtained within the Fermi and the
homogeneously-charged-sphere models, as first suggested in \cite{franosch:91}. We
have checked that a wide class of more general models for the nuclear-charge
distribution yields results well within the error bars obtained in this way.

The next 3 lines contain the corrections due to the one-, two-, and three-photon
exchange, respectively. QED values for the two-photon exchange correction were taken
from our previous evaluations \cite{yerokhin:00:prl,artemyev:03}.
The results for the three-photon exchange correction were obtained in this work
within many-body perturbation theory (MBPT), with retaining the Breit interaction to
the first order only. For uranium, we report good agreement with the previous
evaluations of this effect \cite{zherebtsov:00,andreev:01}. The error ascribed to
the three-photon exchange correction is due to incompleteness of the MBPT treatment.
It was estimated by calculating the third-order MBPT contribution with two and more
Breit interactions for each state involved in the transition, adding these
contributions quadratically, and multiplying the result by a conservative factor of
2.

The entry labeled ``One-loop QED'' represents the sum of the first-order self-energy
and vacuum-polarization corrections calculated on hydrogenic wave functions
\cite{mohr:98}. The next line (``Screened QED'') contains the results for the
screened self-energy and vacuum-polarization corrections
\cite{artemyev:99,yerokhin:99:sescr,yerokhin:05:OS}. The uncertainty ascribed to
this entry is the estimation of higher-order screening effects; it was obtained by
multiplying the correction by the ratio of the entries ``Screened QED'' and
``One-loop QED''. The last two lines contain the values for the relativistic recoil
correction \cite{artemyev:95:pra,artemyev:95:jpb} and the nuclear polarization
correction \cite{plunien:95,nefiodov:96}.

The comparison presented in Table~\ref{tab:3} demonstrates that our total results
agree well within the error bars specified with the experimental data for bismuth
and uranium. The theoretical accuracy is significantly better in the former case,
which is the consequence of the fact that the finite nuclear size effect is smaller
and the nuclear radius is known better. Our result for the $2p_{3/2}$-$2s$
transition in bismuth can also be compared with the value of $2787.96$~eV obtained
by Sapirstein and Cheng \cite{sapirstein:01:lamb}. The difference of 0.16~eV between
the results is mainly due to the two-loop Lamb shift contribution (0.12~eV) which is
not accounted for in Ref.~\cite{sapirstein:01:lamb}.

We conclude that inclusion of the two-loop Lamb shift is necessary for adequate
interpretation of the experimental result in the case of bismuth, whereas for
uranium the two-loop Lamb shift is significantly screened by the uncertainty due to
the nuclear charge distribution. Comparison of the theoretical and experimental
results for bismuth yields the first identification of the two-loop QED effects in
the region of strong binding field, which is a step toward the test of the
strong-field regime of bound-state QED at the two-loop level.

Valuable discussions with U.~Jentschura and K.~Pachucki are gratefully acknowledged.
This work was supported by INTAS YS grant No.~03-55-1442, by the "Dynasty"
foundation, and by RFBR grants No.~04-02-17574 and No.~06-02-04007. The computation
was partly performed on the CINES and IDRIS French national computer centers.
Laboratoire Kastler Brossel is Unit{\'e} Mixte de Recherche du CNRS n$^{\circ}$
8552.

%\bibliographystyle{d:/papers/bibtex/phaip30}
%\bibliography{d:/papers/bibtex/hfst}

\end{document}